\documentclass[prd, twocolumn, aastex]{revtex4-1}
\pdfoutput=1 
\usepackage{amsmath,amstext}
\usepackage[T1]{fontenc}
\usepackage{apjfonts} 
\usepackage{graphicx}
\usepackage{natbib}
\usepackage{xcolor}
\usepackage{hyperref}

\usepackage{aasjournal}

\begin{document}

\title{Reconstructing the weak lensing magnification distribution of Type Ia supernovae}

\author{Zhongxu Zhai$^*$}

\author{Yun Wang}

\affiliation{IPAC, California Institute of Technology, Mail Code 314-6, 1200 E. California Blvd., Pasadena, CA 91125}

\email{zhai@ipac.caltech.edu}

\begin{abstract}
Weak lensing of Type Ia supernovae (SNe Ia) is a systematic uncertainty in the use of SNe Ia as standard candles, as well as an independent cosmological probe, if the corresponding magnification distribution can be extracted from data. We study the peak brightness distribution of SNe Ia in the Pantheon sample, and find that the high $z$ sub-sample shows distinct weak lensing signatures compared to the low $z$ subsample: a long tail at the bright end due to high magnifications and a shift of the peak brightness toward the faint end, consistent with findings from earlier work. We have developed a technique to reconstruct the weak lensing magnification distribution of SNe Ia, $p(\mu)$, from the measured SN Ia flux distribution, and applied it to the Pantheon sample. We find that $p(\mu)$ can be reconstructed at a significance better than 2$\sigma$ for the subsample of SNe Ia at $z>0.7$ (124 SNe Ia), and at a lower significance for the SNe Ia at $z>0.9$ (49 SNe Ia), due to the small number of SNe Ia at high redshifts. The large number of $z>1$ SNe Ia from future surveys will enable the use of $p(\mu)$ reconstructed from SNe Ia as an independent cosmological probe. 

\end{abstract}


\maketitle

\section{Introduction}

Type Ia supernovae (SNe Ia) as standard candles play an important role in modern cosmology. The luminosity distance-redshift relation obtained through their observation provides a powerful probe of the expansion of the universe, and led to the discovery of cosmic acceleration \citep{Riess_1998, Perlmutter_1999}. Over the past decades, various surveys have collectively observed thousands of SNe Ia \citep{Riess_1999, Riess_2004, Astier_2006, Miknaitis_2007, Conley_2011, Frieman_2008, Suzuki_2012, Rest_2014, Graur_2014}, and they provide strong constraints on the matter-energy components in the universe \citep{Amanullah_2010, Betoule_2014, Scolnic_2018}. In order to achieve accurate and precise cosmological constraints, accurate modeling of supernovae with comprehensive examination of the systematic uncertainties is of critical importance. The effects induced by weak gravitational lensing of SNe Ia is one of the main systematic uncertainties, and its impact increases with redshift \citep{Wambsganss_1997, Holz_1998, Valageas_2000a, Wang_2002}. 
On one hand, the bias in cosmological inference due to weak lensing can be minimized/removed using flux-averaging \citep{Wang_2000,Wang_2004}.
On the other hand, the weak lensing magnification of SNe Ia contains important information of the distribution of matter in the universe \citep{Wang_1999}. Future surveys will target thousands of SNe Ia at $z>1$ \citep{Spergel_2015, LSST-sciece-book}. The accurate modeling of the weak lensing effect with this high statistics can significantly improve our understanding of the properties of dark matter and dark energy.

Due to the inhomogeneous distribution of matter in the universe, the light emitted from SNe Ia is bent along the line of sight to the observer. This effect leads to the magnification of the brightness of the observed SNe Ia and affect the scatter from the mean brightness. Compared with the intrinsic brightness distribution of SNe Ia, this weak lensing signature is subdominant. However, for high redshift objects, this effect is not negligible since the light can experience more bending before reaching the observer. This weak lensing effect can be expressed in terms of a probability function of magnification, see e.g. \cite{Valageas_2000a, Wang_2002, Vale_2003, Takahashi_2011} and references therein. 
The resulting distribution of SN Ia brightness is thus a convolution of this magnification distribution and the intrinsic distribution of brightness. The latest SNe Ia sample has a compilation of more than one thousand data points, this enables us to perform a thorough analysis to explore the possible signals of weak lensing in current observation. This extends the earlier investigation in \cite{Wang_2005_wl}.

The systematic caused by weak lensing can turn into signal when our modeling and observation are sufficiently improved \citep{Dodelson_2006, Marra_2013, Quartin_2014}. The weak lensing signature observed in the SN Ia data contains information of the underlying distribution of matter which depends on cosmology. A method that can extract this information can provide useful information to constrain the matter distribution and cosmological parameters. 
We present a methodology for reconstructing the weak lensing magnification distribution from the observed peak flux distribution of SNe Ia, and apply it to the Pantheon sample compiled by Scolnic et al. (2018) \cite{Scolnic_2018}, to demonstrate the feasibility of this approach.

Our paper is organized as follows: we present the modeling of weak lensing signature in the SNe Ia observation in Section 2, as well as the results from the application to the Pantheon sample. We measure the scatter of the intrinsic brightness of SNe Ia in Section 3, and reconstruct the weak lensing magnification distribution of SNe Ia in Section 4. Section 5 presents our discussion and conclusion.

\section{Weak lensing signature}\label{sec:signature}

The derivation of the effect of weak lensing on the magnification of supernovae has been discussed with details in \cite{Bernardeau_1997, Kaiser_1998, Valageas_2000a, Valageas_2000b, Wang_2002}. Here we follow the pioneering work in Wang (2005) \cite{Wang_2005_wl} and briefly describe the weak lensing signature in the type Ia supernova observations.

Due to the intervening matter and structure, the light received by the observer is bent and this can modify the observed brightness of SNe Ia. The observed flux from a SNe Ia can be written as
\begin{equation}
    f=\mu L_{\text{int}},
\end{equation}
where $L_{\text{int}}$ is the intrinsic brightness of the SNe Ia, and $\mu$ is the magnification due to lensing, which can be modeled by a universal probability distribution function based on the measured matter power spectrum \citep{Wang_2002}. The two variables $L_{\text{int}}$ and $\mu$ are statistically independent, therefore the distribution of their product $f$ can be modeled explicitly with the probability distribution function (PDF) of each variables. The resulting distribution can be written as 
\begin{equation}\label{eq:pdf_f}
    p(f)=\int_{0}^{L_{\text{int}}^{\text{max}}}\frac{dL_{\text{int}}}{L_{\text{int}}}g(L_{\text{int}})p\left(\frac{f}{L_{\text{int}}}\right),
\end{equation}
where $p(\mu)$ is the PDF of the magnification of SNe Ia, and $g(L_{\text{int}})$ is the PDF of the intrinsic brightness of SNe. The upper limit of the integration $L_{\text{int}}^{\text{max}}=f/\mu_{\text{min}}$, resulting from the requirement $\mu=f/L_{\text{int}}\geq\mu_{\text{min}}$, where $\mu_{\text{min}}$ is the minimum value of the magnification due to lensing and can be computed for a given cosmological model. Without prior knowledge for the distribution of the intrinsic brightness of SNe Ia, we follow \cite{Wang_2005_wl} and assume that $g(L_{\text{int}})$ is a Gaussian distribution with unit mean and dispersion $\sigma$. The value of $\sigma$ can be well estimated with a large sample of SNe Ia at low redshift, however we will show that this quantity can also be measured as a byproduct in our weak lensing analysis.

\begin{figure*}[htbp]
\begin{center}
\includegraphics[width=18cm]{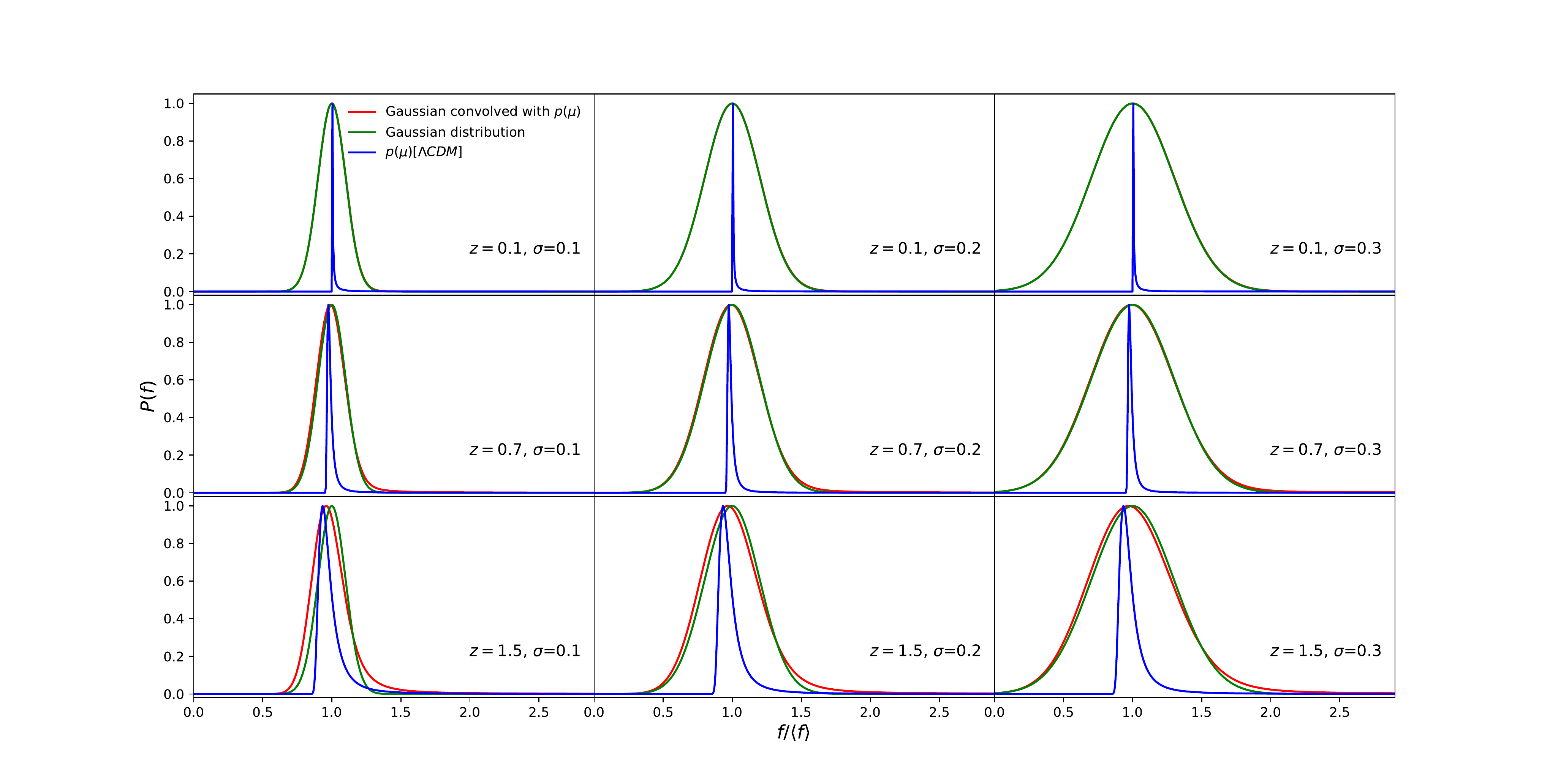}
\caption{Prediction of the observed flux distributions of SNe Ia for magnification distribution $p(\mu)$ and intrinsic brightness distribution $g(L_{\text{int}})$ with different widths and redshifts. The cosmological model is obtained through flux-averaging method and the width of $g(L_{\text{int}})$ is expressed in unit of the mean flux. The distributions are normalized to have peak value equal to 1.}
\label{fig:pdf}
\end{center}
\end{figure*}

In this paper, we measure $p(\mu)$ by using the universal probability distribution function (UPDF) of weak lensing amplification as presented in \cite{Wang_1999, Wang_2005_wl}. Note that $p(\mu)$ can also be computed with analytic method or using cosmological N-body simulations, see e.g. \cite{Valageas_2000a, Barber_2000, Premadi_2001, Vale_2003, Wambsganss_1997, Yoo_2008, Takahashi_2011} and references therein. In this UPDF based framework, we first calculate the minimum convergence as \cite{Wang_2005mincov}
\begin{equation}\label{eq:kappa}
    \hat{\kappa}_{\text{min}}(z)=-\frac{3}{2}\frac{\Omega_{m}(1+z)}{cH_{0}^{-1}}\int_{0}^{z}dz'\frac{(1+z')^{2}}{E(z')}\frac{r(z')}{r(z)}[\lambda(z)-\lambda(z')],
\end{equation}
where $r(z)$ is the comoving distance in a smooth universe,
\begin{equation}
    E(z)\equiv\sqrt{\Omega_{m}(1+z)^3+\Omega_{\Lambda}+\Omega_{k}(1+z)^2}
\end{equation}
is the dimensionless Hubble parameter in a $\Lambda$CDM cosmology with $\Omega_{x}$ denoting the matter-energy fraction of the corresponding component $x$ (here `m' refers to matter, `$\Lambda$' refers to cosmological constant, and `k' refers to curvature contribution). The affine parameter 
\begin{equation}
    \lambda(z) = cH_{0}^{-1}\int_{0}^{z}\frac{dz'}{(1+z')^{2}E(z')}.
\end{equation}
The minimum of magnification $\mu_{\text{min}}$ is related to the minimum of convergence through $\mu_{\text{min}}=1/(1-\hat{\kappa}_{\text{min}})^2$. This relation can be derived in terms of angular diameter distances as detailed in Chapter 4 of \cite{wang2009dark}. Based on the numerical simulation, the data of $p(\mu)$ is converted to a modified UPDF of the reduced convergence $\eta$ \citep{Wang_2005_wl},
\begin{equation} \label{eq:P_eta}
    P(\eta)=\frac{1}{1+\eta^2}\exp\left[ -\left(\frac{\eta-\eta_{\text{peak}}}{\omega\eta^{q}}\right)^{2} \right],
\end{equation}
where 
\begin{equation} \label{eq:P_eta2}
    \eta=1+\frac{\mu-1}{|\mu_{\text{min}}-1|}.
\end{equation}
The parameters in this formula $\eta_{\text{peak}}, \omega, q$ are functions of the variance of $\eta$, $\xi_{\eta}$ which absorbs all the cosmological dependence. For an arbitrary cosmological model, one can compute $\xi_{\eta}$ as \citep{Valageas_2000a}
\begin{equation}
    \xi_{\eta}=\int_{0}^{\chi_{s}}d\chi\left(\frac{w}{F_{s}}\right)I_{\mu}(\chi),
\end{equation}
with
\begin{eqnarray}
    &&F_{s}=\int_{0}^{\chi_{s}}d\chi w(\chi, \chi_{s}), \quad I_{\mu}=\pi\int_{0}^{\infty}\frac{dk}{k}\frac{\Delta^2(k,z)}{k}W^2(Dk\theta_{0}), \nonumber \\
    && \Delta^2(k,z)=4\pi k^3P_{m}(k,z), \quad W(Dk\theta_{0})=\frac{2J_{1}(Dk\theta_{0})}{Dk\theta_{0}} \\
\end{eqnarray}
where $P_{m}(k, z)$ is the matter power spectrum at redshift $z$ with wavenumber $k$, $\theta_{0}$ is the smoothing angle \citep{Valageas_2000b}, and $J_{1}$ is the Bessel function of order 1. The other quantities depending on the distance measure in the universe can be calculated as
\begin{eqnarray}
    && w(\chi, \chi_{s}) = \frac{H_{0}^2}{c^{2}}\frac{D(\chi)D(\chi_{s}-\chi)}{D(\chi_{s})}(1+z)  \nonumber \\
    && D(\chi) = \frac{cH_{0}^{-1}}{\sqrt{|\Omega_{k}|}}\text{sinn}\left(\sqrt{|\Omega_{k}|}\chi\right), \nonumber \\
    && \chi = \int_{0}^{z}\frac{cH_{0}^{-1}dz'}{E(z')},
\end{eqnarray}
where ``sinn" is defined as sinh if $\Omega_{k}>0$, sin if $\Omega_{k}<0$. If $\Omega_{k}=0$, both sinn and $\Omega_{k}$ disappear.

\begin{figure}[htbp]
\begin{center}
\includegraphics[width=8.5cm]{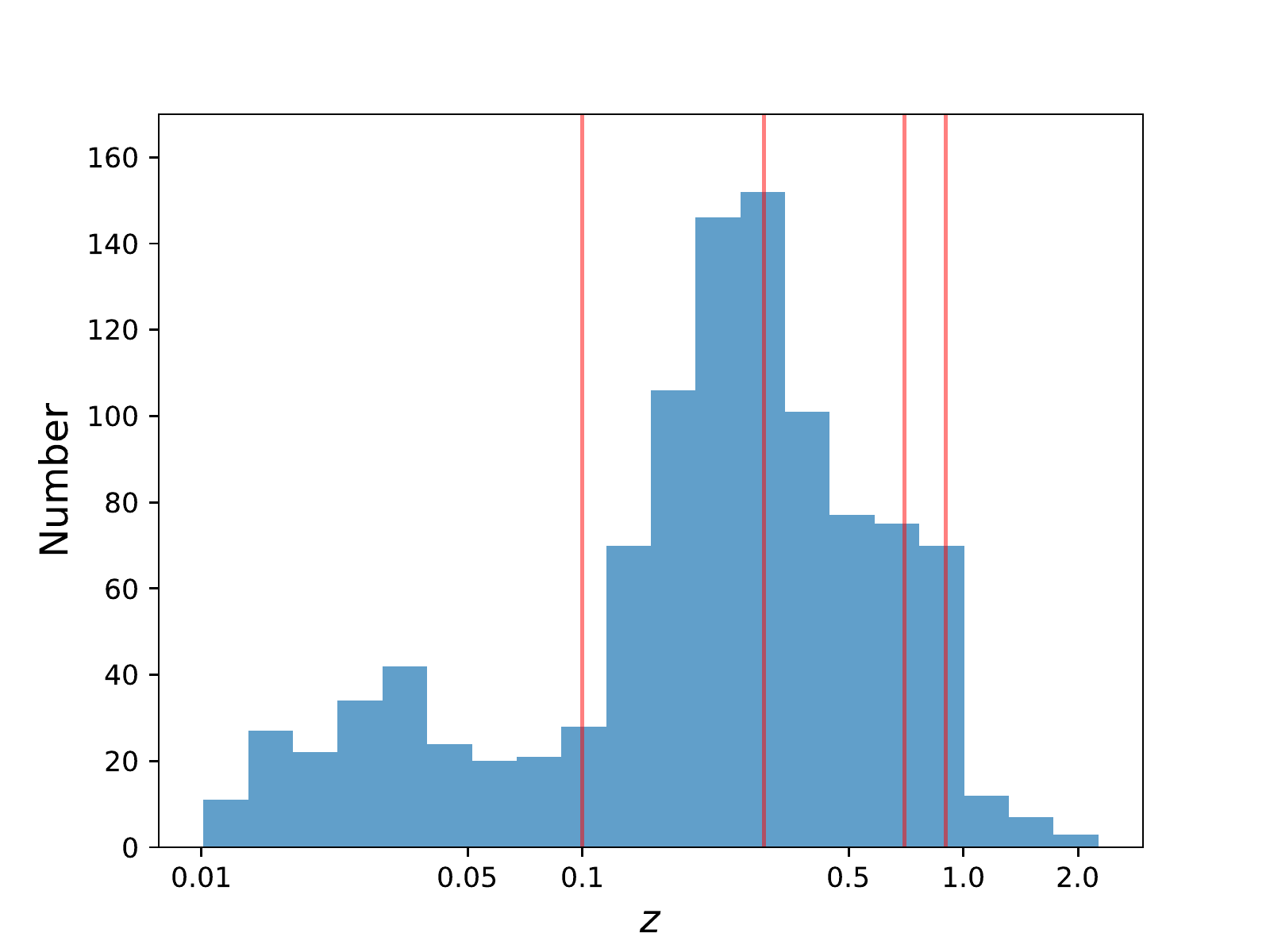}
\caption{The redshift distribution of the Pantheon SNe Ia sample. The vertical red lines correspond to the redshift cuts we use in the analysis: z=0.1, 0.3, 0.7, 0.9.}
\label{fig:z_hist}
\end{center}
\end{figure}

\begin{figure}[htbp]
\begin{center}
\includegraphics[width=8.5cm]{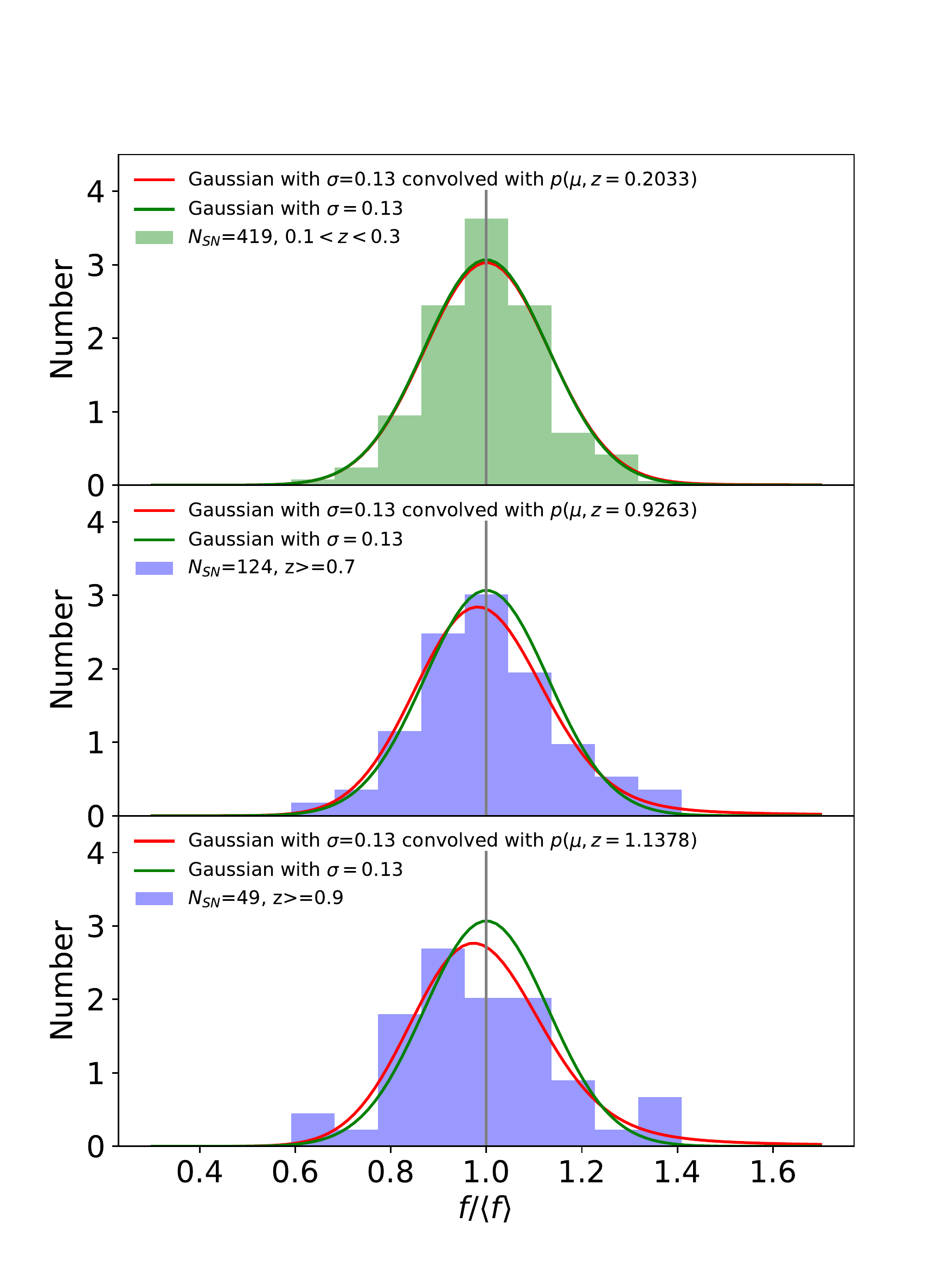}
\caption{The flux distribution of the low-z and high-z SNe Ia samples. The redshift ranges and number of SNe Ia are shown in the legend for each panel. The solid curves in each panel are calculated assuming $\sigma=0.13$ in the Gaussian distribution of the intrinsic brightness distribution, which can be estimated by a simple likelihood analysis. The vertical grey line shows the unity of the mean flux. $p(\mu)$ is calculated at the effective redshift of each sample. The low-z sample has distribution consistent with Gaussian which can be used to anchor the mean flux for both low-z and the high-z samples. The $z>0.7$ sample has better statistics than the $z>0.9$ sample, due to larger size. It presents characteristic signatures of weak lensing on the SNe Ia observations as explained in the context.}
\label{fig:signature}
\end{center}
\end{figure}

In our analysis, we adopt the fitting formula provided by \cite{Wang_2002} and further improved in \cite{wang2009dark} to calculate $p(\mu)$ and convolve it with the intrinsic brightness distribution to obtain the observed flux distribution (see Eq[\ref{eq:pdf_f}]). Figure \ref{fig:pdf} presents the prediction of the observed flux distributions of SNe Ia for magnification distribution $p(\mu)$ at various redshifts and with different widths of the intrinsic brightness distribution $g(L_{\text{int}})$. The cosmological model adopted is obtained through flux-averaging method with the latest Pantheon SNe Ia sample \citep{Zhai_2018}. The result presents clear signatures of the weak lensing effect of SNe Ia data which are consistent with earlier investigations: a non-Gaussian tail at the bright end due to high magnifications, and a shift of the peak towards the faint end due to de-magnification since the Universe is mostly empty. We also find that these signatures become more significant at high redshift and with narrower intrinsic brightness distribution. This is due to the fact that the light emitted from high redshift SNe Ia can experience more bending before reaching the observer and thus result in stronger lensing effects.

Next we explore this weak lensing effect in the current SNe Ia data. We use the Pantheon sample compiled from the full set of spectroscopically confirmed Pan-STARRS1 (PS1) SNe Ia with the observation from CfA1-4, CSP, PS1, SDSS, SNLS and Hubble Space Telescope (HST) SN surveys \citep{Scolnic_2018}. This dataset consists of 1048 SNe Ia in the redshift range 0.01<z<2.3. Figure \ref{fig:z_hist} displays the redshift distribution of this sample, with some cuts used in the analysis shown as vertical lines.  Compared with the datasets analyzed in earlier investigations \citep{Wang_2005_wl}, this enlarged catalog can have significantly improved statistics. In order to isolate the weak lensing signal in the SNe Ia data, we first separate the data into several redshift bins. The low redshift data with z<0.1 are not considered in the analysis since they are significantly affected by the peculiar velocities. As we present in Figure \ref{fig:pdf}, the low redshift SNe Ia do not have detectable weak lensing signature, therefore we isolate the data with $0.1<z<0.3$ to calculate the mean flux and use this value to normalize the SNe Ia at higher redshift. This can enable a meaningful and self-consistent comparison of the high-z and low-z samples. 

We present the resulting distribution of the SNe Ia flux in Figure \ref{fig:signature}. The top panel shows that the low-z sample is consistent with a Gaussian distribution with $\sigma=0.13$ as we may expect. For comparison, the Gaussian distribution and the predicted distribution of SNe Ia flux from convolution of a Gaussian distribution and $p(\mu)$ are also shown. Here $p(\mu)$ is calculated by Eq.(\ref{eq:P_eta}) and (\ref{eq:P_eta2}) at the effective redshift of the sample. The middle panel and the bottom panel show the results for high-z samples with two redshift cuts $z>0.7$ and $z>0.9$ respectively. We note that the high-z sample with $z>0.7$ shows clear signatures of weak lensing: a high magnification at the bright end and a demagnification shift of the peak toward the faint end. This finding is consistent with the earlier study in \citep{Wang_2005_wl} but with better statistics due to the fact that the size of this high-z sample is improved by a factor of two. We also presents result with even higher redshift cut $z>0.9$ in the bottom panel. It implies similar pattern as the $z>0.7$ sample (124 SNe Ia), but with more noise. Therefore we will focus on the $z>0.7$ sample in the following analysis and just briefly present the result from the $z>0.9$ sample (49 SNe Ia) as it is more strongly dominated by shot noise. The curves in Figure \ref{fig:pdf} are obtained assuming $\sigma=0.13$ in the Gaussian distribution of the intrinsic brightness. We will show that this parameter can be estimated from a simple likelihood anlaysis in the next section.

\section{Constraint on the dispersion of the intrinsic brightness}

\begin{figure}[htbp]
\begin{center}
\includegraphics[width=8.5cm]{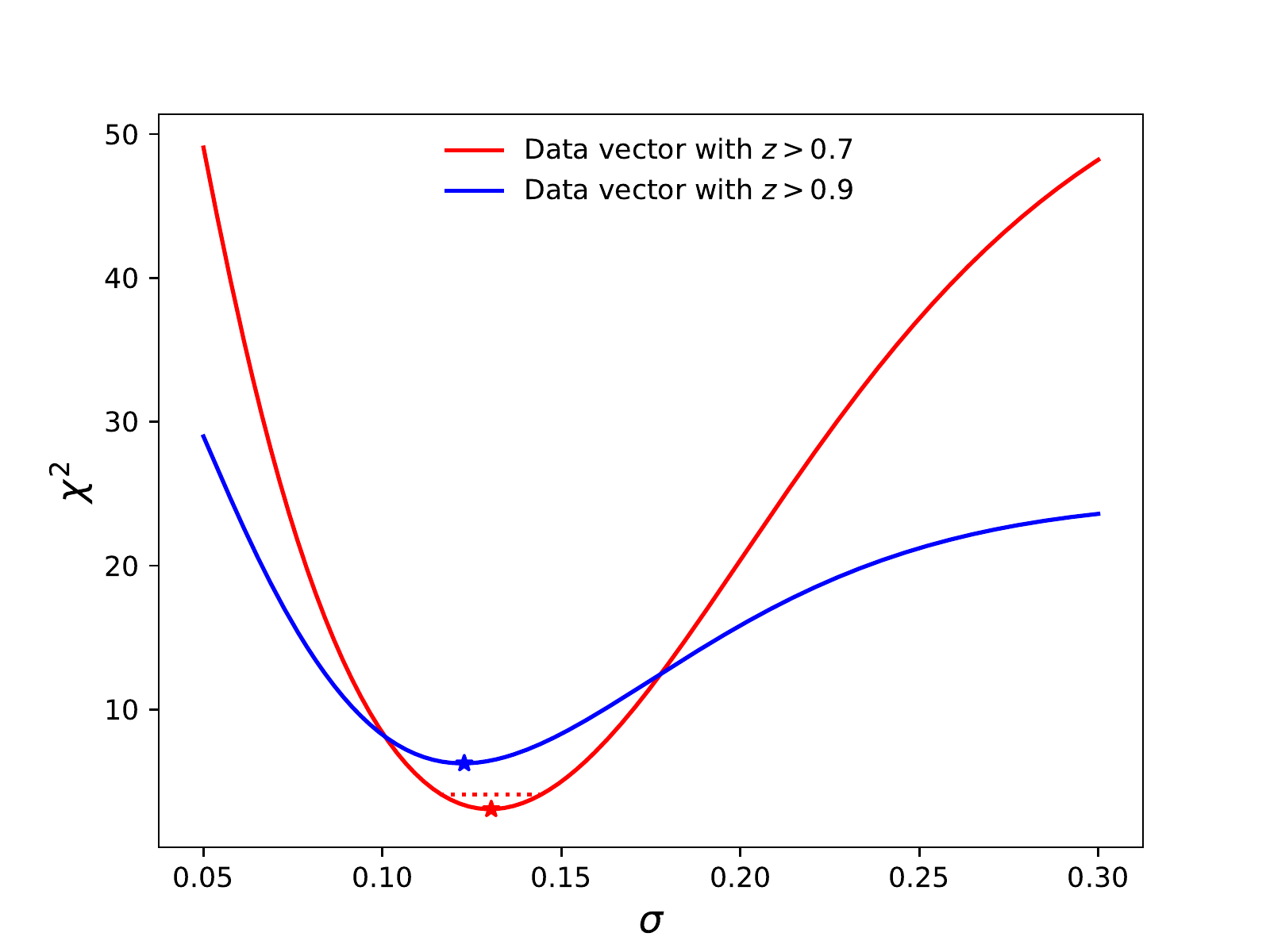}
\caption{Constraints on $\sigma$: the dispersion of the intrinsic brightness distribution of SNe Ia, for two different Pantheon subsamples: $z>0.7$ (red) and $z>0.9$ (blue). The stars denote the best fit value, and the horizontal dotted line for the $z>0.7$ sample corresponds to $\Delta\chi^2=1$ than the best $\chi^2$. The results of these two constraints are consistent with each other at $1\sigma$ level, both indicating that the intrinsic brightness of SNe Ia has a dispersion of about $13\pm1\%$ in unit of the mean value.}
\label{fig:cons_sigma}
\end{center}
\end{figure}

The observed flux distribution shown in Figure \ref{fig:signature} can be used to compare with the theoretical predictions in Eq (\ref{eq:pdf_f}). In this comparison, we assume a cosmological model from flux averaging method \citep{Wang_2004} and the remaining unknown parameter is $\sigma$ in the intrinsic brightness distribution $g(L_{\text{int}})$ of SNe Ia. Therefore we can construct a naive likelihood function
\begin{equation} \label{eq:chi2}
    \chi^{2}=\sum_{i=0}^{N_{\text{bin}}}\frac{D_{i,obs}-D_{i,pre}}{\sigma_{D,i}},
\end{equation}
where $N_{\text{bin}}$ is the number of bins in $f$ as in Figure \ref{fig:signature}, $D_{i}$ is the number of SNe Ia in the $i-th$ bin, the subscripts ``obs" and ``pre" refer to observation and prediction respectively. Since the observable is the number count of SNe Ia in flux, we assume the uncertainty follows a simple Poisson distribution which gives $\sigma_{D,i}=\sqrt{D_{i,obs}}$. We present the constraints on $\sigma$ with two different redshift cuts of the Pantheon sample in Figure \ref{fig:cons_sigma} by calculating the value of $\chi^{2}$ in a range of $\sigma$. The result shows that the two redshift cuts give best-fit values of $\sigma$ consistent with each other. Both samples show a dispersion of $13\%$ in unit of the mean flux of SNe Ia. It could indicate that a redshift-independent intrinsic scatter is a reasonable assumption, but we should note that this result needs to be verified using much larger future samples with much better statistics. The dispersion of intrinsic brightness of SNe Ia can reflect the underlying physical mechanism. Its accurate measurement can provide information of the physics related to explosion model and galaxy environment, and improve the constraints on the cosmological parameters \citep{Kessler_2013}.

\section{Reconstruction of $p(\mu)$}
\label{sec:pmu}

\begin{figure}[htbp]
\begin{center}
\includegraphics[width=8.5cm]{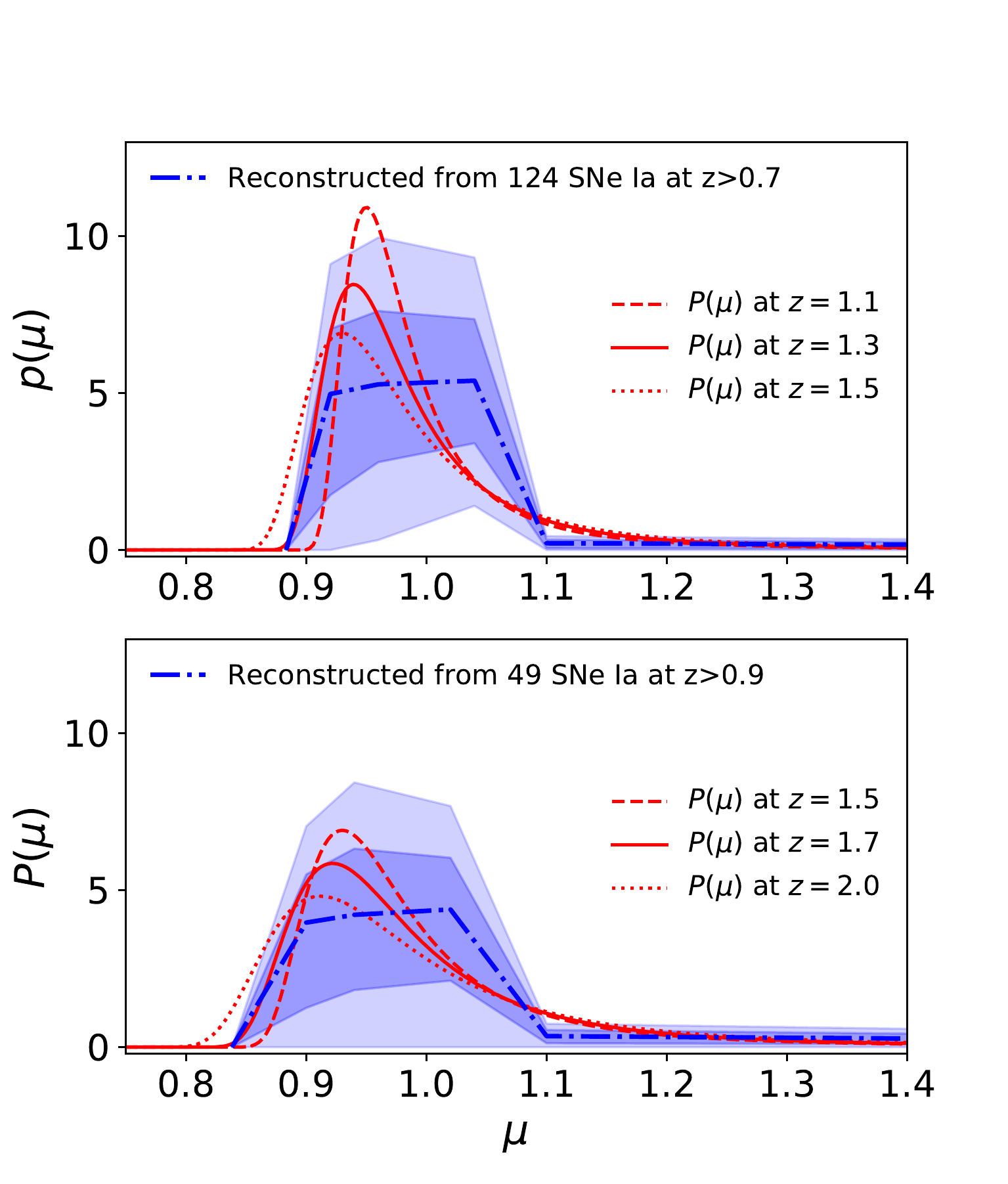}
\caption{Reconstruction of $p(\mu)$ from samples with two different redshift cuts: $Top~panel:$ $z>0.7$; $Bottom~panel:$ $z>0.9$. The blue dot-dashed line is the reconstructed $p(\mu)$ from observation and the shaded area from inner to outer is 1 and 2 $\sigma$ uncertainties. The red lines correspond to the theoretical prediction of $p(\mu)$ from UPDF at various redshifts.}
\label{fig:recons}
\end{center}
\end{figure}

The observed flux distribution of SNe Ia as presented in Figure \ref{fig:signature} is a convolution of the intrinsic distribution and weak lensing magnification. The latter contains important information about the spatial distribution of dark matter and the late time evolution of the universe. Therefore its direct or indirect measurement can be important and challenging, and serves as a new probe to constrain cosmology. In this section, we present a methodology for extracting the measurement of $p(\mu)$ from the observed SN Ia flux distribution, and apply it to current SNe Ia observations.

Our approach is to numerically deconvolve Eq.(\ref{eq:pdf_f}) by parametrizing the weak lensing magnification distribution $p(\mu)$ and assuming a model for the SN Ia peak brightness intrinsic scatter $g(L_{\text{int}})$. Here are the steps in our method:\\
\noindent
(1) Flux-average the SNe Ia to remove/minimize weak lensing effect.\\
\noindent
(2) Derive the bestfit cosmological model using the flux-averaged SNe Ia.\\
\noindent
(3) Derive the SN Ia flux distribution for the high $z$ sub-sample by removing the distance dependence of the SN Ia apparent peak brightness assuming the bestfit cosmological model.\\
\noindent
(4) Parametrize $p(\mu)$ with a set of parameters, \{$p(\mu_i)$\}, for $\mu_{\text{min}} < \mu_i < \mu_{\text{max}}$, where $\mu_{\text{min}}$ is calculated from the minimum of convergence in Eq (\ref{eq:kappa}), and $\mu_{\text{max}}$ is chosen to be large enough such that further increasing its value has negligible effect on the results. Note that $p(\mu)$ is assumed to be zero elsewhere.\\
\noindent
(5) Model the SN Ia peak brightness intrinsic scatter $g(L_{\text{int}})$ as a Gaussian with dispersion $\sigma$.\\
\noindent
(6) Interpolate \{$p(\mu_i)$\} to obtain a model $p(\mu)$, integrate Eq.(\ref{eq:pdf_f}) to obtain predicted flux distribution for the high $z$ sub-sample, and normalize it by the number of SNe Ia in the sub-sample. \\
\noindent
(7) Compute the likelihood function in Eq.(\ref{eq:chi2}).\\
\noindent
(8) Run a Monte Carlo Markov Chain (MCMC) test with the \texttt{emcee} toolkit \citep{Foreman-Mackey_2013} to obtain constraints on the parameters $\{p(\mu_i), \sigma\}$.

The current SN Ia data do not allow a detailed reconstruction of $p(\mu)$. For measuring $p(\mu_i)>0$, we have chosen
$\mu_i=(0.92, 0.96, 1.04, 1.1)$ for the $z>0.7$ sample, and $\mu_i=(0.90, 0.94, 1.02, 1.1)$ for the $z>0.9$ sample respectively. We used the same number of parameters for the two samples, but a wider range in $\mu_i$ for the higher $z$ sample as its $p(\mu)$ is expected to have a broader distribution.

To demonstrate the feasibility of our approach and verify consistency, we have derived the intrinsic scatter in SN Ia peak luminosity separately (see the previous section), instead of estimating it in a joint analysis with $\{p(\mu_i)\}$ as in Step (8) above.
We use linear interpolation in this demonstration for simplicity.
We will explore the joint estimation of $\{p(\mu_i), \sigma\}$ and more sophisticated interpolation schemes in future work.

We present this reconstructed $p(\mu)$ in Figure \ref{fig:recons} for sub-samples with two different redshift cuts, $z>0.7$ (124 SNe Ia), and $z>0.9$ (49 SNe Ia). The red lines show the prediction of $p(\mu)$ based on UPDF (Section \ref{sec:signature}) at various redshifts. The reconstruction from the Pantheon sample is shown as dot-dashed blue line with shaded area. There are two noticeable features in the result: first, the resulting $p(\mu)$ has a peak shape around $\mu< \sim 1.0$, consistent with the UPDF prediction. The significance of the reconstructed $p(\mu)$ is higher than $2\sigma$ for the $z>0.7$ sub-sample, indicating a positive detection of the weak lensing magnification in SNe Ia observations. Second, the reconstructed $p(\mu)$ is broader than the theoretical prediction at the mean redshift of the sample. This is partly due to the fact that the light from the highest redshift supernovae in the sample can experience more bending and result in significant weak lensing magnification. Compared with the limited number of SNe Ia with $z>0.7$ or $z>0.9$, their contribution to the flux distribution is not negligible. The predictions of $p(\mu)$ based on UPDF at higher redshifts show that the high-z SNe Ia can dramatically change the overall shape of $p(\mu)$. The higher redshift subsample has enhanced weak lensing signatures but degraded detectability due to the reduced statistics (see bottom panel of Fig.\ref{fig:recons}).

Fig.\ref{fig:recons} demonstrates that the methodology presented in this section can be used to extract $p(\mu)$ from SN Ia data.
We expect that it will lead to detailed measurement of $p(\mu)$ for high $z$ SNe Ia when applied to sufficiently large samples of SNe Ia at $z>1$.

\section{Discussion and conclusion}

We have presented a detection of the weak lensing magnification in the SNe Ia observations. It extends the pioneering work in \cite{Wang_2005_wl} to derive the the distribution of SNe Ia brightness. By the use of the latest Pantheon SNe Ia sample, we find consistent weak lensing signatures in the flux distribution but with better statistics: a high magnification tail at the bright end and a shift of peak magnification to $\mu<1$ toward the faint end. Our analysis uses the low redshift sample to find the mean flux and the results for two high redshift samples are shown for comparison. 

The observed flux distribution of SNe Ia is a result of convolution between a intrinsic brightness distribution and weak lensing magnification. We assume the intrinsic distribution is Gaussian distributed with unit mean and unknown dispersion $\sigma$. With measurement for the observed flux distribution, we construct and perform a simple likelihood analysis and obtain the constraint on the dispersion. For two samples with different redshift cuts, we find $\sigma=0.13$ in unit of the mean flux. The results of the two samples are consistent with each other and doesn't present significant redshift dependence. This type of measurement could reveal the physical mechanism of SNe Ia explosion and may contain information about the galaxies that host SNe Ia. 

We have developed a methodology to reconstruct the weak lensing magnification of SNe Ia $p(\mu)$ from their observed peak flux distribution (see Sec.\ref{sec:pmu}).
Our method is straightforward and assumes that $p(\mu)$ is an interpolation at certain values of $\mu$ and the measurements are determined from a MCMC analysis. We applied our approach to the Pantheon sample of SNe Ia, and reconstructed $p(\mu)$ for two sub-samples: $z>0.7$ (124 SNe Ia), and $z>0.9$ (49 SNe Ia). The significance of the reconstructed $p(\mu)$ is higher than $2\sigma$ for the $z>0.7$ sub-sample, and lower for the $z>0.9$ sub-sample due to its smaller size.

We have assumed that selection effects have been accurately modeled and corrected in the Pantheon sample by Scolnic et al. (2018) \cite{Scolnic_2018}. Since this is a complex and challenging issue, we will examine it in detail using realistically simulated data in future work.

As a direct probe of dark matter and dark matter halos, weak lensing provides an important means to study their properties. The reconstruction of $p(\mu)$ from the SNe Ia observation is an independent measure of the underlying distribution of matter. This in turn implies that we can use this measurement to constrain cosmology \citep{Wang_2000,Dodelson_2006,Wang_2005}. In this case, the SNe Ia will not only be a geometrical probe, but also provide constraint on the matter distribution in the universe. Even though the measurement of $p(\mu)$ can be noisy, it provides an independent cross-check in cosmological constraints and can help break the degeneracy of the cosmological parameters. 
The methodology presented in this paper can be applied to the thousands of SNe Ia at $z>1$ that WFIRST will observe \citep{Spergel_2015}, and yield unique information to improve our understanding of the universe.

\acknowledgements
We thank Dan Scolnic for helpful discussions. This work was supported in part by NASA grant 15-WFIRST15-0008 Cosmology with the High Latitude Survey WFIRST Science Investigation Team (SIT).

\bibliography{DE_GoF,software}

\begin{thebibliography}{41}
\expandafter\ifx\csname natexlab\endcsname\relax\def\natexlab#1{#1}\fi
\expandafter\ifx\csname bibnamefont\endcsname\relax
  \def\bibnamefont#1{#1}\fi
\expandafter\ifx\csname bibfnamefont\endcsname\relax
  \def\bibfnamefont#1{#1}\fi
\expandafter\ifx\csname citenamefont\endcsname\relax
  \def\citenamefont#1{#1}\fi
\expandafter\ifx\csname url\endcsname\relax
  \def\url#1{\texttt{#1}}\fi
\expandafter\ifx\csname urlprefix\endcsname\relax\def\urlprefix{URL }\fi
\providecommand{\bibinfo}[2]{#2}
\providecommand{\eprint}[2][]{\url{#2}}

\bibitem[{\citenamefont{{Riess} et~al.}(1998)\citenamefont{{Riess},
  {Filippenko}, {Challis}, {Clocchiatti}, {Diercks}, {Garnavich}, {Gilliland},
  {Hogan}, {Jha}, {Kirshner} et~al.}}]{Riess_1998}
\bibinfo{author}{\bibfnamefont{A.~G.} \bibnamefont{{Riess}}},
  \bibinfo{author}{\bibfnamefont{A.~V.} \bibnamefont{{Filippenko}}},
  \bibinfo{author}{\bibfnamefont{P.}~\bibnamefont{{Challis}}},
  \bibinfo{author}{\bibfnamefont{A.}~\bibnamefont{{Clocchiatti}}},
  \bibinfo{author}{\bibfnamefont{A.}~\bibnamefont{{Diercks}}},
  \bibinfo{author}{\bibfnamefont{P.~M.} \bibnamefont{{Garnavich}}},
  \bibinfo{author}{\bibfnamefont{R.~L.} \bibnamefont{{Gilliland}}},
  \bibinfo{author}{\bibfnamefont{C.~J.} \bibnamefont{{Hogan}}},
  \bibinfo{author}{\bibfnamefont{S.}~\bibnamefont{{Jha}}},
  \bibinfo{author}{\bibfnamefont{R.~P.} \bibnamefont{{Kirshner}}},
  \bibnamefont{et~al.}, \bibinfo{journal}{Astron. J.}
  \textbf{\bibinfo{volume}{116}}, \bibinfo{pages}{1009} (\bibinfo{year}{1998}),
  \eprint{astro-ph/9805201}.

\bibitem[{\citenamefont{{Perlmutter} et~al.}(1999)\citenamefont{{Perlmutter},
  {Aldering}, {Goldhaber}, {Knop}, {Nugent}, {Castro}, {Deustua}, {Fabbro},
  {Goobar}, {Groom} et~al.}}]{Perlmutter_1999}
\bibinfo{author}{\bibfnamefont{S.}~\bibnamefont{{Perlmutter}}},
  \bibinfo{author}{\bibfnamefont{G.}~\bibnamefont{{Aldering}}},
  \bibinfo{author}{\bibfnamefont{G.}~\bibnamefont{{Goldhaber}}},
  \bibinfo{author}{\bibfnamefont{R.~A.} \bibnamefont{{Knop}}},
  \bibinfo{author}{\bibfnamefont{P.}~\bibnamefont{{Nugent}}},
  \bibinfo{author}{\bibfnamefont{P.~G.} \bibnamefont{{Castro}}},
  \bibinfo{author}{\bibfnamefont{S.}~\bibnamefont{{Deustua}}},
  \bibinfo{author}{\bibfnamefont{S.}~\bibnamefont{{Fabbro}}},
  \bibinfo{author}{\bibfnamefont{A.}~\bibnamefont{{Goobar}}},
  \bibinfo{author}{\bibfnamefont{D.~E.} \bibnamefont{{Groom}}},
  \bibnamefont{et~al.}, \bibinfo{journal}{\apj} \textbf{\bibinfo{volume}{517}},
  \bibinfo{pages}{565} (\bibinfo{year}{1999}), \eprint{astro-ph/9812133}.

\bibitem[{\citenamefont{{Riess} et~al.}(1999)\citenamefont{{Riess}, {Kirshner},
  {Schmidt}, {Jha}, {Challis}, {Garnavich}, {Esin}, {Carpenter}, {Grashius},
  {Schild} et~al.}}]{Riess_1999}
\bibinfo{author}{\bibfnamefont{A.~G.} \bibnamefont{{Riess}}},
  \bibinfo{author}{\bibfnamefont{R.~P.} \bibnamefont{{Kirshner}}},
  \bibinfo{author}{\bibfnamefont{B.~P.} \bibnamefont{{Schmidt}}},
  \bibinfo{author}{\bibfnamefont{S.}~\bibnamefont{{Jha}}},
  \bibinfo{author}{\bibfnamefont{P.}~\bibnamefont{{Challis}}},
  \bibinfo{author}{\bibfnamefont{P.~M.} \bibnamefont{{Garnavich}}},
  \bibinfo{author}{\bibfnamefont{A.~A.} \bibnamefont{{Esin}}},
  \bibinfo{author}{\bibfnamefont{C.}~\bibnamefont{{Carpenter}}},
  \bibinfo{author}{\bibfnamefont{R.}~\bibnamefont{{Grashius}}},
  \bibinfo{author}{\bibfnamefont{R.~E.} \bibnamefont{{Schild}}},
  \bibnamefont{et~al.}, \bibinfo{journal}{\aj} \textbf{\bibinfo{volume}{117}},
  \bibinfo{pages}{707} (\bibinfo{year}{1999}), \eprint{astro-ph/9810291}.

\bibitem[{\citenamefont{{Riess} et~al.}(2004)\citenamefont{{Riess}, {Strolger},
  {Tonry}, {Casertano}, {Ferguson}, {Mobasher}, {Challis}, {Filippenko}, {Jha},
  {Li} et~al.}}]{Riess_2004}
\bibinfo{author}{\bibfnamefont{A.~G.} \bibnamefont{{Riess}}},
  \bibinfo{author}{\bibfnamefont{L.-G.} \bibnamefont{{Strolger}}},
  \bibinfo{author}{\bibfnamefont{J.}~\bibnamefont{{Tonry}}},
  \bibinfo{author}{\bibfnamefont{S.}~\bibnamefont{{Casertano}}},
  \bibinfo{author}{\bibfnamefont{H.~C.} \bibnamefont{{Ferguson}}},
  \bibinfo{author}{\bibfnamefont{B.}~\bibnamefont{{Mobasher}}},
  \bibinfo{author}{\bibfnamefont{P.}~\bibnamefont{{Challis}}},
  \bibinfo{author}{\bibfnamefont{A.~V.} \bibnamefont{{Filippenko}}},
  \bibinfo{author}{\bibfnamefont{S.}~\bibnamefont{{Jha}}},
  \bibinfo{author}{\bibfnamefont{W.}~\bibnamefont{{Li}}}, \bibnamefont{et~al.},
  \bibinfo{journal}{\apj} \textbf{\bibinfo{volume}{607}}, \bibinfo{pages}{665}
  (\bibinfo{year}{2004}), \eprint{astro-ph/0402512}.

\bibitem[{\citenamefont{{Astier} et~al.}(2006)\citenamefont{{Astier}, {Guy},
  {Regnault}, {Pain}, {Aubourg}, {Balam}, {Basa}, {Carlberg}, {Fabbro},
  {Fouchez} et~al.}}]{Astier_2006}
\bibinfo{author}{\bibfnamefont{P.}~\bibnamefont{{Astier}}},
  \bibinfo{author}{\bibfnamefont{J.}~\bibnamefont{{Guy}}},
  \bibinfo{author}{\bibfnamefont{N.}~\bibnamefont{{Regnault}}},
  \bibinfo{author}{\bibfnamefont{R.}~\bibnamefont{{Pain}}},
  \bibinfo{author}{\bibfnamefont{E.}~\bibnamefont{{Aubourg}}},
  \bibinfo{author}{\bibfnamefont{D.}~\bibnamefont{{Balam}}},
  \bibinfo{author}{\bibfnamefont{S.}~\bibnamefont{{Basa}}},
  \bibinfo{author}{\bibfnamefont{R.~G.} \bibnamefont{{Carlberg}}},
  \bibinfo{author}{\bibfnamefont{S.}~\bibnamefont{{Fabbro}}},
  \bibinfo{author}{\bibfnamefont{D.}~\bibnamefont{{Fouchez}}},
  \bibnamefont{et~al.}, \bibinfo{journal}{\aap} \textbf{\bibinfo{volume}{447}},
  \bibinfo{pages}{31} (\bibinfo{year}{2006}), \eprint{astro-ph/0510447}.

\bibitem[{\citenamefont{{Miknaitis} et~al.}(2007)\citenamefont{{Miknaitis},
  {Pignata}, {Rest}, {Wood-Vasey}, {Blondin}, {Challis}, {Smith}, {Stubbs},
  {Suntzeff}, {Foley} et~al.}}]{Miknaitis_2007}
\bibinfo{author}{\bibfnamefont{G.}~\bibnamefont{{Miknaitis}}},
  \bibinfo{author}{\bibfnamefont{G.}~\bibnamefont{{Pignata}}},
  \bibinfo{author}{\bibfnamefont{A.}~\bibnamefont{{Rest}}},
  \bibinfo{author}{\bibfnamefont{W.~M.} \bibnamefont{{Wood-Vasey}}},
  \bibinfo{author}{\bibfnamefont{S.}~\bibnamefont{{Blondin}}},
  \bibinfo{author}{\bibfnamefont{P.}~\bibnamefont{{Challis}}},
  \bibinfo{author}{\bibfnamefont{R.~C.} \bibnamefont{{Smith}}},
  \bibinfo{author}{\bibfnamefont{C.~W.} \bibnamefont{{Stubbs}}},
  \bibinfo{author}{\bibfnamefont{N.~B.} \bibnamefont{{Suntzeff}}},
  \bibinfo{author}{\bibfnamefont{R.~J.} \bibnamefont{{Foley}}},
  \bibnamefont{et~al.}, \bibinfo{journal}{\apj} \textbf{\bibinfo{volume}{666}},
  \bibinfo{pages}{674} (\bibinfo{year}{2007}), \eprint{astro-ph/0701043}.

\bibitem[{\citenamefont{{Conley} et~al.}(2011)\citenamefont{{Conley}, {Guy},
  {Sullivan}, {Regnault}, {Astier}, {Balland}, {Basa}, {Carlberg}, {Fouchez},
  {Hardin} et~al.}}]{Conley_2011}
\bibinfo{author}{\bibfnamefont{A.}~\bibnamefont{{Conley}}},
  \bibinfo{author}{\bibfnamefont{J.}~\bibnamefont{{Guy}}},
  \bibinfo{author}{\bibfnamefont{M.}~\bibnamefont{{Sullivan}}},
  \bibinfo{author}{\bibfnamefont{N.}~\bibnamefont{{Regnault}}},
  \bibinfo{author}{\bibfnamefont{P.}~\bibnamefont{{Astier}}},
  \bibinfo{author}{\bibfnamefont{C.}~\bibnamefont{{Balland}}},
  \bibinfo{author}{\bibfnamefont{S.}~\bibnamefont{{Basa}}},
  \bibinfo{author}{\bibfnamefont{R.~G.} \bibnamefont{{Carlberg}}},
  \bibinfo{author}{\bibfnamefont{D.}~\bibnamefont{{Fouchez}}},
  \bibinfo{author}{\bibfnamefont{D.}~\bibnamefont{{Hardin}}},
  \bibnamefont{et~al.}, \bibinfo{journal}{\apjs}
  \textbf{\bibinfo{volume}{192}}, \bibinfo{eid}{1} (\bibinfo{year}{2011}),
  \eprint{1104.1443}.

\bibitem[{\citenamefont{{Frieman} et~al.}(2008)\citenamefont{{Frieman},
  {Bassett}, {Becker}, {Choi}, {Cinabro}, {DeJongh}, {Depoy}, {Dilday}, {Doi},
  {Garnavich} et~al.}}]{Frieman_2008}
\bibinfo{author}{\bibfnamefont{J.~A.} \bibnamefont{{Frieman}}},
  \bibinfo{author}{\bibfnamefont{B.}~\bibnamefont{{Bassett}}},
  \bibinfo{author}{\bibfnamefont{A.}~\bibnamefont{{Becker}}},
  \bibinfo{author}{\bibfnamefont{C.}~\bibnamefont{{Choi}}},
  \bibinfo{author}{\bibfnamefont{D.}~\bibnamefont{{Cinabro}}},
  \bibinfo{author}{\bibfnamefont{F.}~\bibnamefont{{DeJongh}}},
  \bibinfo{author}{\bibfnamefont{D.~L.} \bibnamefont{{Depoy}}},
  \bibinfo{author}{\bibfnamefont{B.}~\bibnamefont{{Dilday}}},
  \bibinfo{author}{\bibfnamefont{M.}~\bibnamefont{{Doi}}},
  \bibinfo{author}{\bibfnamefont{P.~M.} \bibnamefont{{Garnavich}}},
  \bibnamefont{et~al.}, \bibinfo{journal}{\aj} \textbf{\bibinfo{volume}{135}},
  \bibinfo{pages}{338} (\bibinfo{year}{2008}), \eprint{0708.2749}.

\bibitem[{\citenamefont{{Suzuki} et~al.}(2012)\citenamefont{{Suzuki}, {Rubin},
  {Lidman}, {Aldering}, {Amanullah}, {Barbary}, {Barrientos}, {Botyanszki},
  {Brodwin}, {Connolly} et~al.}}]{Suzuki_2012}
\bibinfo{author}{\bibfnamefont{N.}~\bibnamefont{{Suzuki}}},
  \bibinfo{author}{\bibfnamefont{D.}~\bibnamefont{{Rubin}}},
  \bibinfo{author}{\bibfnamefont{C.}~\bibnamefont{{Lidman}}},
  \bibinfo{author}{\bibfnamefont{G.}~\bibnamefont{{Aldering}}},
  \bibinfo{author}{\bibfnamefont{R.}~\bibnamefont{{Amanullah}}},
  \bibinfo{author}{\bibfnamefont{K.}~\bibnamefont{{Barbary}}},
  \bibinfo{author}{\bibfnamefont{L.~F.} \bibnamefont{{Barrientos}}},
  \bibinfo{author}{\bibfnamefont{J.}~\bibnamefont{{Botyanszki}}},
  \bibinfo{author}{\bibfnamefont{M.}~\bibnamefont{{Brodwin}}},
  \bibinfo{author}{\bibfnamefont{N.}~\bibnamefont{{Connolly}}},
  \bibnamefont{et~al.}, \bibinfo{journal}{\apj} \textbf{\bibinfo{volume}{746}},
  \bibinfo{eid}{85} (\bibinfo{year}{2012}), \eprint{1105.3470}.

\bibitem[{\citenamefont{{Rest} et~al.}(2014)\citenamefont{{Rest}, {Scolnic},
  {Foley}, {Huber}, {Chornock}, {Narayan}, {Tonry}, {Berger}, {Soderberg},
  {Stubbs} et~al.}}]{Rest_2014}
\bibinfo{author}{\bibfnamefont{A.}~\bibnamefont{{Rest}}},
  \bibinfo{author}{\bibfnamefont{D.}~\bibnamefont{{Scolnic}}},
  \bibinfo{author}{\bibfnamefont{R.~J.} \bibnamefont{{Foley}}},
  \bibinfo{author}{\bibfnamefont{M.~E.} \bibnamefont{{Huber}}},
  \bibinfo{author}{\bibfnamefont{R.}~\bibnamefont{{Chornock}}},
  \bibinfo{author}{\bibfnamefont{G.}~\bibnamefont{{Narayan}}},
  \bibinfo{author}{\bibfnamefont{J.~L.} \bibnamefont{{Tonry}}},
  \bibinfo{author}{\bibfnamefont{E.}~\bibnamefont{{Berger}}},
  \bibinfo{author}{\bibfnamefont{A.~M.} \bibnamefont{{Soderberg}}},
  \bibinfo{author}{\bibfnamefont{C.~W.} \bibnamefont{{Stubbs}}},
  \bibnamefont{et~al.}, \bibinfo{journal}{\apj} \textbf{\bibinfo{volume}{795}},
  \bibinfo{eid}{44} (\bibinfo{year}{2014}), \eprint{1310.3828}.

\bibitem[{\citenamefont{{Graur} et~al.}(2014)\citenamefont{{Graur}, {Rodney},
  {Maoz}, {Riess}, {Jha}, {Postman}, {Dahlen}, {Holoien}, {McCully}, {Patel}
  et~al.}}]{Graur_2014}
\bibinfo{author}{\bibfnamefont{O.}~\bibnamefont{{Graur}}},
  \bibinfo{author}{\bibfnamefont{S.~A.} \bibnamefont{{Rodney}}},
  \bibinfo{author}{\bibfnamefont{D.}~\bibnamefont{{Maoz}}},
  \bibinfo{author}{\bibfnamefont{A.~G.} \bibnamefont{{Riess}}},
  \bibinfo{author}{\bibfnamefont{S.~W.} \bibnamefont{{Jha}}},
  \bibinfo{author}{\bibfnamefont{M.}~\bibnamefont{{Postman}}},
  \bibinfo{author}{\bibfnamefont{T.}~\bibnamefont{{Dahlen}}},
  \bibinfo{author}{\bibfnamefont{T.~W.~S.} \bibnamefont{{Holoien}}},
  \bibinfo{author}{\bibfnamefont{C.}~\bibnamefont{{McCully}}},
  \bibinfo{author}{\bibfnamefont{B.}~\bibnamefont{{Patel}}},
  \bibnamefont{et~al.}, \bibinfo{journal}{\apj} \textbf{\bibinfo{volume}{783}},
  \bibinfo{eid}{28} (\bibinfo{year}{2014}), \eprint{1310.3495}.

\bibitem[{\citenamefont{{Amanullah} et~al.}(2010)\citenamefont{{Amanullah},
  {Lidman}, {Rubin}, {Aldering}, {Astier}, {Barbary}, {Burns}, {Conley},
  {Dawson}, {Deustua} et~al.}}]{Amanullah_2010}
\bibinfo{author}{\bibfnamefont{R.}~\bibnamefont{{Amanullah}}},
  \bibinfo{author}{\bibfnamefont{C.}~\bibnamefont{{Lidman}}},
  \bibinfo{author}{\bibfnamefont{D.}~\bibnamefont{{Rubin}}},
  \bibinfo{author}{\bibfnamefont{G.}~\bibnamefont{{Aldering}}},
  \bibinfo{author}{\bibfnamefont{P.}~\bibnamefont{{Astier}}},
  \bibinfo{author}{\bibfnamefont{K.}~\bibnamefont{{Barbary}}},
  \bibinfo{author}{\bibfnamefont{M.~S.} \bibnamefont{{Burns}}},
  \bibinfo{author}{\bibfnamefont{A.}~\bibnamefont{{Conley}}},
  \bibinfo{author}{\bibfnamefont{K.~S.} \bibnamefont{{Dawson}}},
  \bibinfo{author}{\bibfnamefont{S.~E.} \bibnamefont{{Deustua}}},
  \bibnamefont{et~al.}, \bibinfo{journal}{\apj} \textbf{\bibinfo{volume}{716}},
  \bibinfo{pages}{712} (\bibinfo{year}{2010}), \eprint{1004.1711}.

\bibitem[{\citenamefont{{Betoule} et~al.}(2014)\citenamefont{{Betoule},
  {Kessler}, {Guy}, {Mosher}, {Hardin}, {Biswas}, {Astier}, {El-Hage}, {Konig},
  {Kuhlmann} et~al.}}]{Betoule_2014}
\bibinfo{author}{\bibfnamefont{M.}~\bibnamefont{{Betoule}}},
  \bibinfo{author}{\bibfnamefont{R.}~\bibnamefont{{Kessler}}},
  \bibinfo{author}{\bibfnamefont{J.}~\bibnamefont{{Guy}}},
  \bibinfo{author}{\bibfnamefont{J.}~\bibnamefont{{Mosher}}},
  \bibinfo{author}{\bibfnamefont{D.}~\bibnamefont{{Hardin}}},
  \bibinfo{author}{\bibfnamefont{R.}~\bibnamefont{{Biswas}}},
  \bibinfo{author}{\bibfnamefont{P.}~\bibnamefont{{Astier}}},
  \bibinfo{author}{\bibfnamefont{P.}~\bibnamefont{{El-Hage}}},
  \bibinfo{author}{\bibfnamefont{M.}~\bibnamefont{{Konig}}},
  \bibinfo{author}{\bibfnamefont{S.}~\bibnamefont{{Kuhlmann}}},
  \bibnamefont{et~al.}, \bibinfo{journal}{\aap} \textbf{\bibinfo{volume}{568}},
  \bibinfo{eid}{A22} (\bibinfo{year}{2014}), \eprint{1401.4064}.

\bibitem[{\citenamefont{{Scolnic} et~al.}(2018)\citenamefont{{Scolnic},
  {Jones}, {Rest}, {Pan}, {Chornock}, {Foley}, {Huber}, {Kessler}, {Narayan},
  {Riess} et~al.}}]{Scolnic_2018}
\bibinfo{author}{\bibfnamefont{D.~M.} \bibnamefont{{Scolnic}}},
  \bibinfo{author}{\bibfnamefont{D.~O.} \bibnamefont{{Jones}}},
  \bibinfo{author}{\bibfnamefont{A.}~\bibnamefont{{Rest}}},
  \bibinfo{author}{\bibfnamefont{Y.~C.} \bibnamefont{{Pan}}},
  \bibinfo{author}{\bibfnamefont{R.}~\bibnamefont{{Chornock}}},
  \bibinfo{author}{\bibfnamefont{R.~J.} \bibnamefont{{Foley}}},
  \bibinfo{author}{\bibfnamefont{M.~E.} \bibnamefont{{Huber}}},
  \bibinfo{author}{\bibfnamefont{R.}~\bibnamefont{{Kessler}}},
  \bibinfo{author}{\bibfnamefont{G.}~\bibnamefont{{Narayan}}},
  \bibinfo{author}{\bibfnamefont{A.~G.} \bibnamefont{{Riess}}},
  \bibnamefont{et~al.}, \bibinfo{journal}{\apj} \textbf{\bibinfo{volume}{859}},
  \bibinfo{eid}{101} (\bibinfo{year}{2018}), \eprint{1710.00845}.

\bibitem[{\citenamefont{{Wambsganss} et~al.}(1997)\citenamefont{{Wambsganss},
  {Cen}, {Xu}, and {Ostriker}}}]{Wambsganss_1997}
\bibinfo{author}{\bibfnamefont{J.}~\bibnamefont{{Wambsganss}}},
  \bibinfo{author}{\bibfnamefont{R.}~\bibnamefont{{Cen}}},
  \bibinfo{author}{\bibfnamefont{G.}~\bibnamefont{{Xu}}}, \bibnamefont{and}
  \bibinfo{author}{\bibfnamefont{J.~P.} \bibnamefont{{Ostriker}}},
  \bibinfo{journal}{\apjl} \textbf{\bibinfo{volume}{475}}, \bibinfo{pages}{L81}
  (\bibinfo{year}{1997}).

\bibitem[{\citenamefont{{Holz} and {Wald}}(1998)}]{Holz_1998}
\bibinfo{author}{\bibfnamefont{D.~E.} \bibnamefont{{Holz}}} \bibnamefont{and}
  \bibinfo{author}{\bibfnamefont{R.~M.} \bibnamefont{{Wald}}},
  \bibinfo{journal}{\prd} \textbf{\bibinfo{volume}{58}}, \bibinfo{eid}{063501}
  (\bibinfo{year}{1998}), \eprint{astro-ph/9708036}.

\bibitem[{\citenamefont{{Valageas}}(2000{\natexlab{a}})}]{Valageas_2000a}
\bibinfo{author}{\bibfnamefont{P.}~\bibnamefont{{Valageas}}},
  \bibinfo{journal}{\aap} \textbf{\bibinfo{volume}{354}}, \bibinfo{pages}{767}
  (\bibinfo{year}{2000}{\natexlab{a}}), \eprint{astro-ph/9904300}.

\bibitem[{\citenamefont{{Wang} et~al.}(2002)\citenamefont{{Wang}, {Holz}, and
  {Munshi}}}]{Wang_2002}
\bibinfo{author}{\bibfnamefont{Y.}~\bibnamefont{{Wang}}},
  \bibinfo{author}{\bibfnamefont{D.~E.} \bibnamefont{{Holz}}},
  \bibnamefont{and} \bibinfo{author}{\bibfnamefont{D.}~\bibnamefont{{Munshi}}},
  \bibinfo{journal}{\apj} \textbf{\bibinfo{volume}{572}}, \bibinfo{pages}{L15}
  (\bibinfo{year}{2002}), \eprint{astro-ph/0204169}.

\bibitem[{\citenamefont{{Wang}}(2000)}]{Wang_2000}
\bibinfo{author}{\bibfnamefont{Y.}~\bibnamefont{{Wang}}},
  \bibinfo{journal}{\apj} \textbf{\bibinfo{volume}{536}}, \bibinfo{pages}{531}
  (\bibinfo{year}{2000}), \eprint{astro-ph/9907405}.

\bibitem[{\citenamefont{{Wang} and {Mukherjee}}(2004)}]{Wang_2004}
\bibinfo{author}{\bibfnamefont{Y.}~\bibnamefont{{Wang}}} \bibnamefont{and}
  \bibinfo{author}{\bibfnamefont{P.}~\bibnamefont{{Mukherjee}}},
  \bibinfo{journal}{\apj} \textbf{\bibinfo{volume}{606}}, \bibinfo{pages}{654}
  (\bibinfo{year}{2004}), \eprint{astro-ph/0312192}.

\bibitem[{\citenamefont{{Wang}}(1999)}]{Wang_1999}
\bibinfo{author}{\bibfnamefont{Y.}~\bibnamefont{{Wang}}},
  \bibinfo{journal}{\apj} \textbf{\bibinfo{volume}{525}}, \bibinfo{pages}{651}
  (\bibinfo{year}{1999}), \eprint{astro-ph/9901212}.

\bibitem[{\citenamefont{{Spergel} et~al.}(2015)\citenamefont{{Spergel},
  {Gehrels}, {Baltay}, {Bennett}, {Breckinridge}, {Donahue}, {Dressler},
  {Gaudi}, {Greene}, {Guyon} et~al.}}]{Spergel_2015}
\bibinfo{author}{\bibfnamefont{D.}~\bibnamefont{{Spergel}}},
  \bibinfo{author}{\bibfnamefont{N.}~\bibnamefont{{Gehrels}}},
  \bibinfo{author}{\bibfnamefont{C.}~\bibnamefont{{Baltay}}},
  \bibinfo{author}{\bibfnamefont{D.}~\bibnamefont{{Bennett}}},
  \bibinfo{author}{\bibfnamefont{J.}~\bibnamefont{{Breckinridge}}},
  \bibinfo{author}{\bibfnamefont{M.}~\bibnamefont{{Donahue}}},
  \bibinfo{author}{\bibfnamefont{A.}~\bibnamefont{{Dressler}}},
  \bibinfo{author}{\bibfnamefont{B.~S.} \bibnamefont{{Gaudi}}},
  \bibinfo{author}{\bibfnamefont{T.}~\bibnamefont{{Greene}}},
  \bibinfo{author}{\bibfnamefont{O.}~\bibnamefont{{Guyon}}},
  \bibnamefont{et~al.}, \bibinfo{journal}{ArXiv e-prints}
  (\bibinfo{year}{2015}), \eprint{1503.03757}.

\bibitem[{\citenamefont{{LSST Science Collaboration}
  et~al.}(2009)\citenamefont{{LSST Science Collaboration}, {Abell}, {Allison},
  {Anderson}, {Andrew}, {Angel}, {Armus}, {Arnett}, {Asztalos}, {Axelrod}
  et~al.}}]{LSST-sciece-book}
\bibinfo{author}{\bibnamefont{{LSST Science Collaboration}}},
  \bibinfo{author}{\bibfnamefont{P.~A.} \bibnamefont{{Abell}}},
  \bibinfo{author}{\bibfnamefont{J.}~\bibnamefont{{Allison}}},
  \bibinfo{author}{\bibfnamefont{S.~F.} \bibnamefont{{Anderson}}},
  \bibinfo{author}{\bibfnamefont{J.~R.} \bibnamefont{{Andrew}}},
  \bibinfo{author}{\bibfnamefont{J.~R.~P.} \bibnamefont{{Angel}}},
  \bibinfo{author}{\bibfnamefont{L.}~\bibnamefont{{Armus}}},
  \bibinfo{author}{\bibfnamefont{D.}~\bibnamefont{{Arnett}}},
  \bibinfo{author}{\bibfnamefont{S.~J.} \bibnamefont{{Asztalos}}},
  \bibinfo{author}{\bibfnamefont{T.~S.} \bibnamefont{{Axelrod}}},
  \bibnamefont{et~al.}, \bibinfo{journal}{ArXiv e-prints}
  (\bibinfo{year}{2009}), \eprint{0912.0201}.

\bibitem[{\citenamefont{{Vale} and {White}}(2003)}]{Vale_2003}
\bibinfo{author}{\bibfnamefont{C.}~\bibnamefont{{Vale}}} \bibnamefont{and}
  \bibinfo{author}{\bibfnamefont{M.}~\bibnamefont{{White}}},
  \bibinfo{journal}{\apj} \textbf{\bibinfo{volume}{592}}, \bibinfo{pages}{699}
  (\bibinfo{year}{2003}), \eprint{astro-ph/0303555}.

\bibitem[{\citenamefont{{Takahashi} et~al.}(2011)\citenamefont{{Takahashi},
  {Oguri}, {Sato}, and {Hamana}}}]{Takahashi_2011}
\bibinfo{author}{\bibfnamefont{R.}~\bibnamefont{{Takahashi}}},
  \bibinfo{author}{\bibfnamefont{M.}~\bibnamefont{{Oguri}}},
  \bibinfo{author}{\bibfnamefont{M.}~\bibnamefont{{Sato}}}, \bibnamefont{and}
  \bibinfo{author}{\bibfnamefont{T.}~\bibnamefont{{Hamana}}},
  \bibinfo{journal}{\apj} \textbf{\bibinfo{volume}{742}}, \bibinfo{eid}{15}
  (\bibinfo{year}{2011}), \eprint{1106.3823}.

\bibitem[{\citenamefont{{Wang}}(2005)}]{Wang_2005_wl}
\bibinfo{author}{\bibfnamefont{Y.}~\bibnamefont{{Wang}}},
  \bibinfo{journal}{\jcap} \textbf{\bibinfo{volume}{3}}, \bibinfo{eid}{005}
  (\bibinfo{year}{2005}), \eprint{astro-ph/0406635}.

\bibitem[{\citenamefont{{Dodelson} and {Vallinotto}}(2006)}]{Dodelson_2006}
\bibinfo{author}{\bibfnamefont{S.}~\bibnamefont{{Dodelson}}} \bibnamefont{and}
  \bibinfo{author}{\bibfnamefont{A.}~\bibnamefont{{Vallinotto}}},
  \bibinfo{journal}{\prd} \textbf{\bibinfo{volume}{74}}, \bibinfo{eid}{063515}
  (\bibinfo{year}{2006}), \eprint{astro-ph/0511086}.

\bibitem[{\citenamefont{{Marra} et~al.}(2013)\citenamefont{{Marra}, {Quartin},
  and {Amendola}}}]{Marra_2013}
\bibinfo{author}{\bibfnamefont{V.}~\bibnamefont{{Marra}}},
  \bibinfo{author}{\bibfnamefont{M.}~\bibnamefont{{Quartin}}},
  \bibnamefont{and}
  \bibinfo{author}{\bibfnamefont{L.}~\bibnamefont{{Amendola}}},
  \bibinfo{journal}{\prd} \textbf{\bibinfo{volume}{88}}, \bibinfo{eid}{063004}
  (\bibinfo{year}{2013}), \eprint{1304.7689}.

\bibitem[{\citenamefont{{Quartin} et~al.}(2014)\citenamefont{{Quartin},
  {Marra}, and {Amendola}}}]{Quartin_2014}
\bibinfo{author}{\bibfnamefont{M.}~\bibnamefont{{Quartin}}},
  \bibinfo{author}{\bibfnamefont{V.}~\bibnamefont{{Marra}}}, \bibnamefont{and}
  \bibinfo{author}{\bibfnamefont{L.}~\bibnamefont{{Amendola}}},
  \bibinfo{journal}{\prd} \textbf{\bibinfo{volume}{89}}, \bibinfo{eid}{023009}
  (\bibinfo{year}{2014}), \eprint{1307.1155}.

\bibitem[{\citenamefont{{Bernardeau} et~al.}(1997)\citenamefont{{Bernardeau},
  {van Waerbeke}, and {Mellier}}}]{Bernardeau_1997}
\bibinfo{author}{\bibfnamefont{F.}~\bibnamefont{{Bernardeau}}},
  \bibinfo{author}{\bibfnamefont{L.}~\bibnamefont{{van Waerbeke}}},
  \bibnamefont{and}
  \bibinfo{author}{\bibfnamefont{Y.}~\bibnamefont{{Mellier}}},
  \bibinfo{journal}{\aap} \textbf{\bibinfo{volume}{322}}, \bibinfo{pages}{1}
  (\bibinfo{year}{1997}), \eprint{astro-ph/9609122}.

\bibitem[{\citenamefont{{Kaiser}}(1998)}]{Kaiser_1998}
\bibinfo{author}{\bibfnamefont{N.}~\bibnamefont{{Kaiser}}},
  \bibinfo{journal}{\apj} \textbf{\bibinfo{volume}{498}}, \bibinfo{pages}{26}
  (\bibinfo{year}{1998}), \eprint{astro-ph/9610120}.

\bibitem[{\citenamefont{{Valageas}}(2000{\natexlab{b}})}]{Valageas_2000b}
\bibinfo{author}{\bibfnamefont{P.}~\bibnamefont{{Valageas}}},
  \bibinfo{journal}{\aap} \textbf{\bibinfo{volume}{356}}, \bibinfo{pages}{771}
  (\bibinfo{year}{2000}{\natexlab{b}}), \eprint{astro-ph/9911336}.

\bibitem[{\citenamefont{{Barber} et~al.}(2000)\citenamefont{{Barber}, {Thomas},
  {Couchman}, and {Fluke}}}]{Barber_2000}
\bibinfo{author}{\bibfnamefont{A.~J.} \bibnamefont{{Barber}}},
  \bibinfo{author}{\bibfnamefont{P.~A.} \bibnamefont{{Thomas}}},
  \bibinfo{author}{\bibfnamefont{H.~M.~P.} \bibnamefont{{Couchman}}},
  \bibnamefont{and} \bibinfo{author}{\bibfnamefont{C.~J.}
  \bibnamefont{{Fluke}}}, \bibinfo{journal}{\mnras}
  \textbf{\bibinfo{volume}{319}}, \bibinfo{pages}{267} (\bibinfo{year}{2000}),
  \eprint{astro-ph/0002437}.

\bibitem[{\citenamefont{{Premadi} et~al.}(2001)\citenamefont{{Premadi},
  {Martel}, {Matzner}, and {Futamase}}}]{Premadi_2001}
\bibinfo{author}{\bibfnamefont{P.}~\bibnamefont{{Premadi}}},
  \bibinfo{author}{\bibfnamefont{H.}~\bibnamefont{{Martel}}},
  \bibinfo{author}{\bibfnamefont{R.}~\bibnamefont{{Matzner}}},
  \bibnamefont{and}
  \bibinfo{author}{\bibfnamefont{T.}~\bibnamefont{{Futamase}}},
  \bibinfo{journal}{\apjs} \textbf{\bibinfo{volume}{135}}, \bibinfo{pages}{7}
  (\bibinfo{year}{2001}), \eprint{astro-ph/0101359}.

\bibitem[{\citenamefont{{Yoo} et~al.}(2008)\citenamefont{{Yoo}, {Ishihara},
  {Nakao}, and {Tagoshi}}}]{Yoo_2008}
\bibinfo{author}{\bibfnamefont{C.}~\bibnamefont{{Yoo}}},
  \bibinfo{author}{\bibfnamefont{H.}~\bibnamefont{{Ishihara}}},
  \bibinfo{author}{\bibfnamefont{K.}~\bibnamefont{{Nakao}}}, \bibnamefont{and}
  \bibinfo{author}{\bibfnamefont{H.}~\bibnamefont{{Tagoshi}}},
  \bibinfo{journal}{Progress of Theoretical Physics}
  \textbf{\bibinfo{volume}{120}}, \bibinfo{pages}{961} (\bibinfo{year}{2008}),
  \eprint{0711.2720}.

\bibitem[{\citenamefont{{Wang} et~al.}(2005)\citenamefont{{Wang}, {Tenbarge},
  and {Fleshman}}}]{Wang_2005mincov}
\bibinfo{author}{\bibfnamefont{Y.}~\bibnamefont{{Wang}}},
  \bibinfo{author}{\bibfnamefont{J.}~\bibnamefont{{Tenbarge}}},
  \bibnamefont{and}
  \bibinfo{author}{\bibfnamefont{B.}~\bibnamefont{{Fleshman}}},
  \bibinfo{journal}{\apj} \textbf{\bibinfo{volume}{624}}, \bibinfo{pages}{46}
  (\bibinfo{year}{2005}), \eprint{astro-ph/0307415}.

\bibitem[{\citenamefont{Wang}(2010)}]{wang2009dark}
\bibinfo{author}{\bibfnamefont{Y.}~\bibnamefont{Wang}},
  \emph{\bibinfo{title}{Dark energy}} (\bibinfo{publisher}{Wiley-VCH},
  \bibinfo{year}{2010}).

\bibitem[{\citenamefont{{Zhai} and {Wang}}(2018)}]{Zhai_2018}
\bibinfo{author}{\bibfnamefont{Z.}~\bibnamefont{{Zhai}}} \bibnamefont{and}
  \bibinfo{author}{\bibfnamefont{Y.}~\bibnamefont{{Wang}}},
  \bibinfo{journal}{arXiv e-prints}  (\bibinfo{year}{2018}),
  \eprint{1811.07425}.

\bibitem[{\citenamefont{{Kessler} et~al.}(2013)\citenamefont{{Kessler}, {Guy},
  {Marriner}, {Betoule}, {Brinkmann}, {Cinabro}, {El-Hage}, {Frieman}, {Jha},
  {Mosher} et~al.}}]{Kessler_2013}
\bibinfo{author}{\bibfnamefont{R.}~\bibnamefont{{Kessler}}},
  \bibinfo{author}{\bibfnamefont{J.}~\bibnamefont{{Guy}}},
  \bibinfo{author}{\bibfnamefont{J.}~\bibnamefont{{Marriner}}},
  \bibinfo{author}{\bibfnamefont{M.}~\bibnamefont{{Betoule}}},
  \bibinfo{author}{\bibfnamefont{J.}~\bibnamefont{{Brinkmann}}},
  \bibinfo{author}{\bibfnamefont{D.}~\bibnamefont{{Cinabro}}},
  \bibinfo{author}{\bibfnamefont{P.}~\bibnamefont{{El-Hage}}},
  \bibinfo{author}{\bibfnamefont{J.~A.} \bibnamefont{{Frieman}}},
  \bibinfo{author}{\bibfnamefont{S.}~\bibnamefont{{Jha}}},
  \bibinfo{author}{\bibfnamefont{J.}~\bibnamefont{{Mosher}}},
  \bibnamefont{et~al.}, \bibinfo{journal}{\apj} \textbf{\bibinfo{volume}{764}},
  \bibinfo{eid}{48} (\bibinfo{year}{2013}), \eprint{1209.2482}.

\bibitem[{\citenamefont{{Foreman-Mackey}
  et~al.}(2013)\citenamefont{{Foreman-Mackey}, {Hogg}, {Lang}, and
  {Goodman}}}]{Foreman-Mackey_2013}
\bibinfo{author}{\bibfnamefont{D.}~\bibnamefont{{Foreman-Mackey}}},
  \bibinfo{author}{\bibfnamefont{D.~W.} \bibnamefont{{Hogg}}},
  \bibinfo{author}{\bibfnamefont{D.}~\bibnamefont{{Lang}}}, \bibnamefont{and}
  \bibinfo{author}{\bibfnamefont{J.}~\bibnamefont{{Goodman}}},
  \bibinfo{journal}{\pasp} \textbf{\bibinfo{volume}{125}}, \bibinfo{pages}{306}
  (\bibinfo{year}{2013}), \eprint{1202.3665}.

\bibitem[{\citenamefont{{Wang} and {Tegmark}}(2005)}]{Wang_2005}
\bibinfo{author}{\bibfnamefont{Y.}~\bibnamefont{{Wang}}} \bibnamefont{and}
  \bibinfo{author}{\bibfnamefont{M.}~\bibnamefont{{Tegmark}}},
  \bibinfo{journal}{\prd} \textbf{\bibinfo{volume}{71}}, \bibinfo{eid}{103513}
  (\bibinfo{year}{2005}), \eprint{astro-ph/0501351}.

\end{thebibliography}

\end{document}